\documentclass[dvips]{article}
\usepackage{supertabular,lscape,epsfig}
\usepackage{amssymb}
\usepackage{amsmath}
\usepackage{rotating}
\DeclareSymbolFont{ppa}{OT1}{ppl}{m}{it}
\DeclareMathSymbol{\vv}{\mathalpha}{ppa}{'166}

\begin{document}

\newcommand{\dd}{\,{\rm d}}
\newcommand{\ie}{{\it i.e.},\,}
\newcommand{\etal}{{\it et al.\ }}
\newcommand{\eg}{{\it e.g.},\,}
\newcommand{\cf}{{\it cf.\ }}
\newcommand{\vs}{{\it vs.\ }}
\newcommand{\zdot}{\makebox[0pt][l]{.}}
\newcommand{\up}[1]{\ifmmode^{\rm #1}\else$^{\rm #1}$\fi}
\newcommand{\dn}[1]{\ifmmode_{\rm #1}\else$_{\rm #1}$\fi}
\newcommand{\upd}{\up{d}}
\newcommand{\uph}{\up{h}}
\newcommand{\upm}{\up{m}}
\newcommand{\ups}{\up{s}}
\newcommand{\arcd}{\ifmmode^{\circ}\else$^{\circ}$\fi}
\newcommand{\arcm}{\ifmmode{'}\else$'$\fi}
\newcommand{\arcs}{\ifmmode{''}\else$''$\fi}
\newcommand{\MS}{{\rm M}\ifmmode_{\odot}\else$_{\odot}$\fi}
\newcommand{\RS}{{\rm R}\ifmmode_{\odot}\else$_{\odot}$\fi}
\newcommand{\LS}{{\rm L}\ifmmode_{\odot}\else$_{\odot}$\fi}
\newcommand{\feh}{\hbox{$ [{\rm Fe}/{\rm H}]$}}

\newcommand{\Abstract}[2]{{\footnotesize\begin{center}ABSTRACT\end{center}
\vspace{1mm}\par#1\par
\noindent
{~}{\it #2}}}

\newcommand{\TabCap}[2]{\begin{center}\parbox[t]{#1}{\begin{center}
  \small {\spaceskip 2pt plus 1pt minus 1pt T a b l e}
  \refstepcounter{table}\thetable \\[2mm]
  \footnotesize #2 \end{center}}\end{center}}

\newcommand{\TableSep}[2]{\begin{table}[p]\vspace{#1}
\TabCap{#2}\end{table}}

\newcommand{\FigCap}[1]{\footnotesize\par\noindent Fig.\  %
  \refstepcounter{figure}\thefigure. #1\par}

\newcommand{\TableFont}{\footnotesize}
\newcommand{\TableFontIt}{\ttit}
\newcommand{\SetTableFont}[1]{\renewcommand{\TableFont}{#1}}

\newcommand{\MakeTable}[4]{\begin{table}[htb]\TabCap{#2}{#3}
  \begin{center} \TableFont \begin{tabular}{#1} #4
  \end{tabular}\end{center}\end{table}}

\newcommand{\MakeTableSep}[4]{\begin{table}[p]\TabCap{#2}{#3}
  \begin{center} \TableFont \begin{tabular}{#1} #4
  \end{tabular}\end{center}\end{table}}

\newenvironment{references}%
{
\footnotesize \frenchspacing
\renewcommand{\thesection}{}
\renewcommand{\in}{{\rm in }}
\renewcommand{\AA}{Astron.\ Astrophys.}
\newcommand{\AAS}{Astron.~Astrophys.~Suppl.~Ser.}
\newcommand{\ApJ}{Astrophys.\ J.}
\newcommand{\ApJS}{Astrophys.\ J.~Suppl.~Ser.}
\newcommand{\ApJL}{Astrophys.\ J.~Letters}
\newcommand{\AJ}{Astron.\ J.}
\newcommand{\IBVS}{IBVS}
\newcommand{\PASJ}{PASJ}
\newcommand{\PASP}{P.A.S.P.}
\newcommand{\Acta}{Acta Astron.}
\newcommand{\MNRAS}{MNRAS}
\renewcommand{\and}{{\rm and }}
\section{{\rm REFERENCES}}
\sloppy \hyphenpenalty10000
\begin{list}{}{\leftmargin1cm\listparindent-1cm
\itemindent\listparindent\parsep0pt\itemsep0pt}}%
{\end{list}\vspace{2mm}}

\def\TYLDA{~}
\newlength{\DW}
\settowidth{\DW}{0}
\newcommand{\dw}{\hspace{\DW}}

\newcommand{\refitem}[5]{\item[]{#1} #2%
\def\REFARG{#3}\ifx\REFARG\TYLDA\else, {\it#3}\fi
\def\REFARG{#4}\ifx\REFARG\TYLDA\else, {\bf#4}\fi
\def\REFARG{#5}\ifx\REFARG\TYLDA\else, {#5}\fi.}

\newcommand{\Section}[1]{\section{#1}}
\newcommand{\Subsection}[1]{\subsection{#1}}
\newcommand{\Acknow}[1]{\par\vspace{5mm}{\bf Acknowledgments.} #1}
\pagestyle{myheadings}

\newfont{\bb}{ptmbi8t at 12pt}
\newcommand{\xrule}{\rule{0pt}{2.5ex}}
\newcommand{\xxrule}{\rule[-1.8ex]{0pt}{4.5ex}}
\def\thefootnote{\fnsymbol{footnote}}

\begin{center}
{\Large\bf New Short-Period Delta Scuti Stars in OGLE-IV Fields toward
the Galactic Bulge\footnote{Based on photometric observations obtained with
the 1.3-m Warsaw telescope at Las Campanas Observatory of the Carnegie
Institution for Science and spectroscopic data collected with the European Southern
Observatory (ESO) NTT telescope at La Silla Observatory under ESO programme 105.20EF.001.}}
\vskip1cm
{\bf
P.~~P~i~e~t~r~u~k~o~w~i~c~z$^1$,~~M.~~R~a~t~a~j~c~z~a~k$^1$,~~I.~~S~o~s~z~y~\'n~s~k~i$^1$,\\
A.~~U~d~a~l~s~k~i$^1$,~~M.~K.~~S~z~y~m~a~\'n~s~k~i$^1$,~~K.~~U~l~a~c~z~y~k$^2$,\\
R.~~P~o~l~e~s~k~i$^1$,~~S.~~K~o~z~{\l}~o~w~s~k~i$^1$,~~J.~~S~k~o~w~r~o~n$^1$,\\
D.~M.~~S~k~o~w~r~o~n$^1$,~~P.~~M~r~\'o~z$^1$,~~K.~~R~y~b~i~c~k~i$^{1,3}$,~~P.~~I~w~a~n~e~k$^1$,\\
M.~~W~r~o~n~a$^1$,~~and~~M.~~G~r~o~m~a~d~z~k~i$^1$\\}
\vskip3mm
{
$^1$ Astronomical Observatory, University of Warsaw, Al. Ujazdowskie 4, 00-478 Warszawa, Poland\\
$^2$ Department of Physics, University of Warwick, Coventry CV4 7AL, UK\\
$^3$ Department of Particle Physics and Astrophysics, Weizmann Institute of Science, Rehovot 76100, Israel\\
}
\end{center}

\Abstract{We report the classification of 24 puzzling short-period variable stars
located in OGLE-IV Galactic bulge fields. The stars are low-amplitude ($<0.05$ mag)
multi-periodic objects with dominant periods between 22 and 54 min whose type could
not have been unambiguously established based on photometry only. A low-resolution
spectroscopic follow-up has shown that all the objects are main sequence A/F-type
stars. Thus, all the variables are $\delta$ Sct-type pulsators.
We have added them to the OGLE-IV Collection of Variable Stars.}

{Catalogs -- Stars: variables: $\delta$~Scuti -- Galaxy: bulge -- Galaxy: disk}


\Section{Introduction}

$\delta$~Sct-type variable stars are pulsating stars with periods
below 0.3~d and $V$-band ($I$-band) amplitudes up to 0.9~mag (0.6~mag).
Most of them are multiple-mode pulsators (Netzel \etal 2022). They have
spectral types from A0 to F6 and luminosity classes from III (giants),
through IV (subgiants), to V (dwarfs). In the Hertzsprung-Russell
diagram, $\delta$~Sct stars reside in the part where the classical
instability strip intersects the main sequence (MS). They can also be found
in the pre-MS and post-MS stages. Space-based observations from the Kepler
satellite revealed that no more than 70 per cent of all stars in that part
pulsate (Murphy \etal 2019). The Kepler data also showed that the majority
of $\delta$~Sct stars are low-amplitude pulsators ($<0.1$~mag)
and only about half of them have amplitudes $>0.001$~mag and appear to be
variable in ground-based observations (Balona and Dziembowski 2011).

The upper bound of 0.3~d (7.2~h) for fundamental-mode $\delta$~Sct stars
was arbitrarily set due to scarcity of Population I classical pulsators with periods
in the range 0.3--1~d. Stars with periods longer than 0.3~d are classified
as classical Cepheids (Soszy\'nski \etal 2008). Until the early 2010s,
shortest periods of the known $\delta$~Sct-type stars were of about 25--30 min
(\eg 25.54 min in $\omega$~Cen V294, 26.25 min in $\omega$~Cen V295 -- Kaluzny \etal 2004).
Holdsworth \etal (2014) searched for high-frequency signals in SuperWASP data
and reported on the detection of over 350 candidates for short-period pulsators
including $\delta$~Sct stars with periods in the range 6--29 min.

Classification of variable stars with periods below 1~h ($\approx0.042$~d)
purely based on photometry is often very difficult due to low amplitudes
and sinusoidal shapes of the light curves. A position in the color-magnitude
diagram is usually insufficient to indicate the true variability type,
particularly in the case of heavily reddened objects located close to the Galactic plane.
There are many types of pulsating stars with periods below 1~h. Among them there
are various types of white dwarfs (\eg ZZ Cet, V777 Her -- Winget and Kepler 2008),
pre-white dwarfs (\eg EL CVn -- Maxted \etal 2013), hot subdwarfs
(\eg V361 Hya, V1093 Her -- Heber \etal 2016), Blue Large-Amplitude Pulsators
(Pietrukowicz \etal 2017, Kupfer \etal 2019), objects from the GW Vir family
including planetary nebulae nuclei variables (PNNV, Quirion \etal 2007),
$\delta$~Sct stars (Rodr\'iguez \etal 2000), and their Population II
analogues, SX Phe stars (Rodr\'iguez and L\'opez-Gonz\'alez 2000).
There is also a group of rapidly oscillating Ap (roAp) variables showing light
variations in the range 4.7--25.8 min and amplitudes of millimagnitudes or
smaller (Kurtz 2022). Recent analysis of TESS observations indicates that these
variables should be treated as normal $\delta$~Sct stars and the term ``roAp''
should be avoided (Balona 2022). Definitive classification of short-period
variables often requires a spectroscopic confirmation. This is the
case of objects presented in the work here.

Searches for short-period variables in 121 OGLE-IV fields covering the inner
Galactic bulge and central part of the Sagittarius Dwarf Spheroidal Galaxy brought
the identification of a total of 10~140 $\delta$~Sct stars with periods in the range
0.033--0.3~d (47.7~min--7.2~h) (Pietrukowicz \etal 2020, Soszy\'nski \etal 2021).
The main criteria used in the classification process were the period value and asymmetry
in the light curve shape. In addition to this set, 157 extremely short variables
with periods in the range 0.01--0.04~d (14.4--57.6~min) were found but could not have
been unambiguously classified. From this sample, 24 stars with sinusoidal variations
stable over years and located in the upper part of the MS in the
color-magnitude diagrams were subject of our spectroscopic follow-up. In Section~2
we provide details on the photometric and spectroscopic observations. Information
on the 24 targets selected for the spectroscopic campaign can be found in Section~3,
while the results are presented in Section~4.


\Section{Observations}

Photometric observations were obtained in the framework of a long-term variability
survey, the Optical Gravitational Lensing Experiment (OGLE), conducted on the 1.3-m
Warsaw telescope at Las Campanas Observatory, Chile. The analyzed in this work data
were collected during the fourth phase of the survey (OGLE-IV) in years 2010--2019.
OGLE monitors the Milky Way's bulge and disk and Magellanic Clouds in searches
for periodic and irregular brightness variations as well as for transient events.
The entire monitored area covers about 3600 deg$^2$. The OGLE-IV camera is a
mosaic of 32 CCD detectors covering a field of 1.4 deg$^2$ at a scale of
0.26 arcsec/pixel. Most of the observations were taken through the Cousins $I$ filter.
The remaining data were taken through the Johnson $V$ filter to secure color information.
Over the whole decade in the area of the inner Galactic bulge, OGLE collected about
181~500 $I$-band frames and 6400 $V$-band frames. The number of $I$-band measurements
per object varies from 60 up to 16~800, depending on the field. The exposure times
were of 100~s in $I$ and 150~s in $V$. In the case of three most crowded fields,
BLG501, BLG505, and BLG512, the cadence of $I$-band observations was as short as
19 min. Time-series photometry was extracted using the Difference Image Analysis
(DIA -- Wo\'zniak 2000), a technique developed especially for dense stellar fields
(Alard and Lupton 1998). Observations of the inner bulge cover a magnitude range
$12.5<I<21.5$. Technical details on the OGLE survey can be found in Udalski \etal (2015).
The OGLE-IV Collection of Variable Stars (OCVS) contains over one million
objects of various variability types: pulsating stars (\eg Soszy\'nski \etal 2019),
spotted stars (\eg Iwanek \etal 2019), eclipsing and ellipsoidal binaries
(\eg Pawlak \etal 2016, Wrona \etal 2022), cataclysmic systems (\eg Mr\'oz \etal 2015).
Despite the huge number of known variable stars, some of newly detected objects
are puzzling and require spectroscopic investigation to properly classify them
and to understand their nature.

Low-resolution spectroscopic observations were conducted under ESO programme 105.20EF.001
in remote visitor mode (due to Covid-19 pandemic restrictions) over four consecutive
nights in gray time, from 2021 June 30/July~1 to July 3/4. The spectra were
collected with the ESO Faint Object Spectrograph and Camera 2 (EFOSC2) mounted
at the Nasmyth focus of the 3.58-m New Technology Telescope (NTT) located at
La Silla Observatory, Chile. Information about the characteristics of this instrument
are provided in Buzzoni \etal (1984). Thin cirrus clouds were present during the
whole observing run but the first half of the third night when they were thick.
All the spectra were taken with grism \#4 covering wavelengths 4085--7520~$\rm \AA$ at
the slit width of $1\zdot\arcs0$ and $2\times2$ binning readout, which provided a spectral
resolution of about 11~$\rm \AA$ at 5000~$\rm \AA$. For most of the objects the slit
was aligned to the parallactic angle, but in some cases the slit was oriented properly
to avoid light from neighboring stars. Exposure times were calculated with the help
of the ESO Exposure Time Calculator\footnote{https://www.eso.org/observing/etc/}
to obtain spectra with a signal-to-noise ratio of 50. For wavelength calibrations,
the He-Ar lamp was used. Three spectrophotometric standard stars were selected
for the flux calibrations. Bias and dome flat-field images were taken at dawn.
All spectra were reduced using the IRAF package\footnote{IRAF was distributed
by the National Optical Astronomy Observatory, USA, which is operated by the
Association of Universities for Research in Astronomy, Inc., under a cooperative
agreement with the National Science Foundation.}. Debiasing, flat-fielding,
wavelength and flux calibrations were performed in the standard way.


\Section{Target objects for the spectroscopic follow-up}

In Table~1, we provide information on the 24 stars selected for our spectroscopic
run. The stars are located toward the Galactic bulge. They have
magnitudes in the range $13.1<I<15.6$ and $14.2<V<17.5$. All the stars are
multi-periodic variables with amplitudes below 0.05 mag in $I$.
We list up to five detected periodicities. The periods were determined
with the help of the TATRY code (Schwarzenberg-Czerny 1996) after subsequent
pre-whitenings.

\begin{sidewaystable}
\centering \caption{\small Photometric parameters of the 24 target stars}
\medskip
{\footnotesize
\begin{tabular}{ccccclllll}
\hline
Name      &     RA(2000)      &     Dec(2000)      & $\langle I \rangle$ & $\langle V \rangle$ &   $P_1$   &   $P_2$  &     $P_3$     &     $P_4$     &     $P_5$ \\
OGLE-     &                   &                    &        [mag]        &        [mag]        &    [d]    &    [d]   &      [d]      &      [d]      &      [d]  \\
BLG-DSCT- & & & & & & & & & \\
\hline
16832 & $17\uph22\upm51\zdot\ups74$ & $-30\arcd16\arcm23\zdot\arcs1$ & 15.53 & 17.49 & 0.033531956(24)  & 0.036224500(35)  & 0.042320454(45)  & -                & -                \\
16833 & $17\uph32\upm20\zdot\ups04$ & $-22\arcd05\arcm49\zdot\arcs1$ & 13.10 & 14.20 & 0.034233043(20)  & 0.039879897(52)  & 0.016609671(15)  & -                & -                \\
16834 & $17\uph35\upm57\zdot\ups82$ & $-26\arcd54\arcm50\zdot\arcs6$ & 14.34 & 15.72 & 0.035165296(11)  & 0.038645168(18)  & 0.040651018(27)  & 0.045402275(52)  & 0.035165937(64)  \\
16835 & $17\uph38\upm33\zdot\ups75$ & $-34\arcd09\arcm26\zdot\arcs4$ & 14.85 & 16.65 & 0.035719287(23)  & 0.039362439(60)  & 0.08913910(39)   & 0.043655649(97)  & 0.04184227(10)   \\
16836 & $17\uph44\upm34\zdot\ups25$ & $-33\arcd17\arcm56\zdot\arcs1$ & 15.28 & 16.76 & 0.035541111(20)  & 0.09749517(25)   & 0.051245588(88)  & -                & -                \\
16837 & $17\uph47\upm50\zdot\ups67$ & $-20\arcd48\arcm46\zdot\arcs4$ & 15.17 & 16.53 & 0.037076495(22)  & 0.046284106(35)  & 0.04357615(11)   & -                & -                \\
16838 & $17\uph50\upm02\zdot\ups34$ & $-33\arcd18\arcm59\zdot\arcs5$ & 14.48 & 15.71 & 0.033961025(13)  & 0.037164189(44)  & 0.036201125(56)  & -                & -                \\
16839 & $17\uph50\upm31\zdot\ups18$ & $-29\arcd28\arcm06\zdot\arcs3$ & 15.06 & 16.30 & 0.025409872(10)  & 0.027650646(25)  & 0.029687043(47)  & 0.032681617(62)  & 0.029680207(58)  \\
16840 & $17\uph51\upm35\zdot\ups11$ & $-28\arcd44\arcm39\zdot\arcs2$ & 15.11 & 16.10 & 0.0278339039(50) & 0.032812117(21)  & 0.06660964(12)   & 0.027834329(29)  & 0.047779888(96)  \\
16841 & $17\uph51\upm46\zdot\ups46$ & $-28\arcd26\arcm05\zdot\arcs3$ & 15.08 & 15.93 & 0.0190090983(40) & 0.0201545952(54) & 0.0201278558(72) & 0.0201484003(80) & 0.0213994311(96) \\
16842 & $17\uph51\upm55\zdot\ups89$ & $-30\arcd17\arcm18\zdot\arcs9$ & 14.66 & 15.79 & 0.0255401151(79) & 0.027133418(24)  & 0.027149755(35)  & 0.028792153(40)  & -                \\
16843 & $17\uph52\upm17\zdot\ups68$ & $-32\arcd42\arcm06\zdot\arcs3$ & 13.82 & 14.84 & 0.028433525(10)  & 0.030864635(36)  & -                & -                & -                \\
16844 & $17\uph54\upm35\zdot\ups55$ & $-29\arcd48\arcm24\zdot\arcs4$ & 15.21 & 16.17 & 0.033516366(13)  & 0.036939131(45)  & 0.04307991(13)   & 0.04435281(15)   & 0.04541343(15)   \\
16845 & $17\uph58\upm16\zdot\ups37$ & $-27\arcd54\arcm43\zdot\arcs8$ & 15.13 & 16.67 & 0.029935884(10)  & 0.034300425(12)  & 0.029934575(21)  & 0.039805145(49)  & 0.037234447(61)  \\
16846 & $17\uph58\upm30\zdot\ups84$ & $-28\arcd25\arcm46\zdot\arcs4$ & 13.86 & 14.62 & 0.0230617618(57) & 0.0198797600(60) & 0.053054509(54)  & 0.027531088(26)  & 0.024992040(24)  \\
16847 & $17\uph58\upm58\zdot\ups94$ & $-28\arcd03\arcm42\zdot\arcs6$ & 14.76 & 15.85 & 0.036425934(12)  & 0.042694792(45)  & 0.048473768(69)  & 0.09609703(28)   & 0.07844828(21)   \\
16848 & $17\uph58\upm59\zdot\ups19$ & $-28\arcd45\arcm44\zdot\arcs2$ & 14.61 & 15.48 & 0.030000956(11)  & 0.031942458(16)  & -                & -                & -                \\
16849 & $18\uph00\upm49\zdot\ups06$ & $-27\arcd35\arcm26\zdot\arcs0$ & 13.82 & 14.59 & 0.0311231198(67) & 0.041716525(14)  & 0.035578305(13)  & 0.033469123(20)  & 0.036362040(27)  \\
16850 & $18\uph02\upm25\zdot\ups70$ & $-27\arcd02\arcm23\zdot\arcs5$ & 14.48 & 15.41 & 0.0158377216(23) & 0.0166989591(23) & 0.0150967282(37) & 0.0167467952(47) & 0.0158803976(48) \\
16851 & $18\uph03\upm25\zdot\ups04$ & $-27\arcd01\arcm41\zdot\arcs7$ & 14.78 & 15.61 & 0.0168271841(31) & 0.0189019685(57) & 0.017930385(11)  & 0.018901708(12)  & 0.019266537(18)  \\
16852 & $18\uph05\upm30\zdot\ups36$ & $-28\arcd28\arcm49\zdot\arcs4$ & 15.57 & 16.68 & 0.036513655(16)  & 0.042300103(34)  & 0.046980198(57)  & 0.036513303(41)  & -                \\
16853 & $18\uph15\upm03\zdot\ups77$ & $-27\arcd33\arcm06\zdot\arcs7$ & 14.37 & 14.92 & 0.034369987(13)  & 0.022029517(14)  & -                & -                & -                \\
16854 & $18\uph16\upm13\zdot\ups66$ & $-26\arcd53\arcm15\zdot\arcs3$ & 13.67 & 14.40 & 0.036765128(19)  & 0.048780768(51)  & 0.039913311(33)  & 0.047063512(55)  & 0.020849953(31)  \\
16855 & $18\uph20\upm37\zdot\ups13$ & $-23\arcd00\arcm08\zdot\arcs0$ & 14.62 & 15.64 & 0.034885725(23)  & 0.037769423(12)  & 0.0361497383(56) & -                & -                \\
\hline
\end{tabular}}
\end{sidewaystable}


\Section{Results from the spectroscopic observations}

\begin{figure}
\centerline{\includegraphics[angle=90,width=105mm]{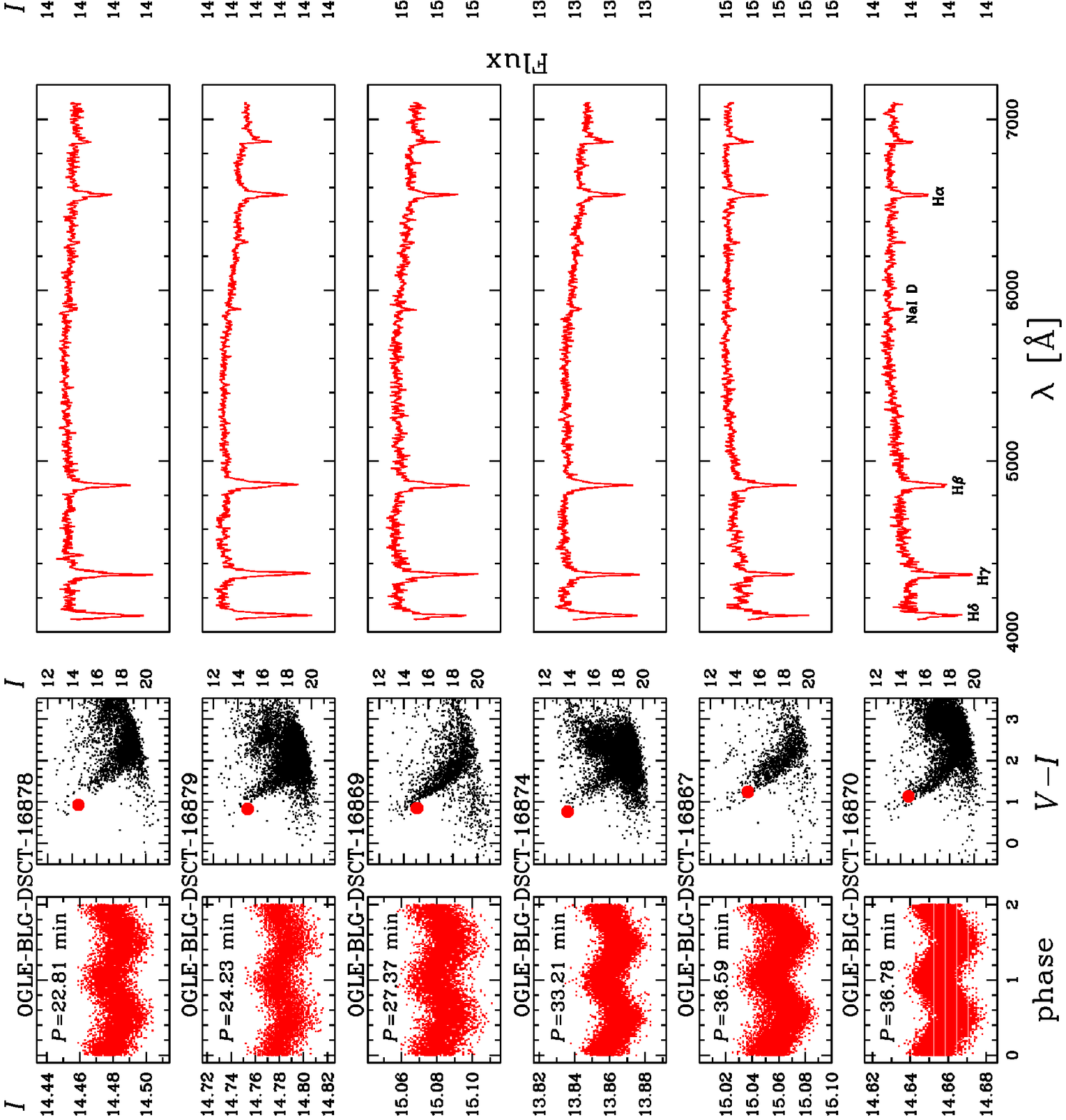}}
\FigCap{Phased OGLE $I$-band light curves (left column), $V-I$ color \vs $I$
magnitude diagrams (middle column), and EFOSC2 low-resolution spectra (right column)
of 12 (out of 24) discovered $\delta$~Sct stars ordered with increasing
pulsation period (provided in the left column).}
\end{figure}

\begin{figure}
\centerline{\includegraphics[angle=90,width=105mm]{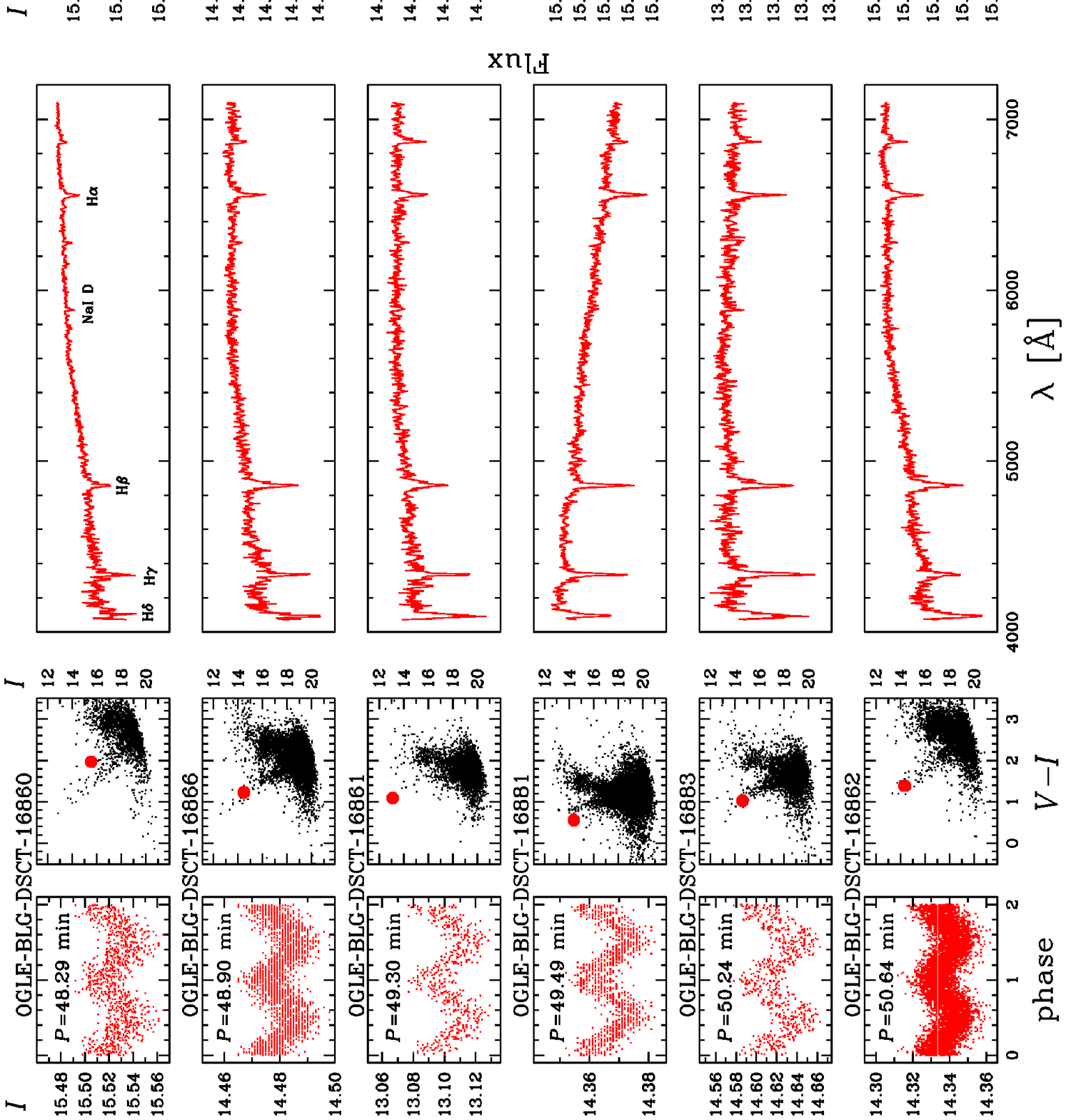}}
\FigCap{Same as Figs.~1 for the remaining $\delta$~Sct stars.}
\end{figure}

In Figs.~1--2, for each star, we present the OGLE-IV $I$-band light curve
folded with the dominant period, a color-magnitude diagram with the position
of the star, and the obtained EFOSC2 spectrum. The spectra of all stars are
dominated by the Balmer series, which indicates type A. In all the stars,
one can also spot the blended sodium doublet Na~I (D1: 5896~$\rm \AA$,
D2: 5890~$\rm \AA$), which is primarily of interstellar origin for this
spectral type.

A detailed analysis of the spectra was carried out with the iSpec package
(Blanco-Cuaresma \etal 2014, Blanco-Cuaresma 2019). To determine atmospheric
parameters, the effective temperature $T_{\rm eff}$ and surface gravity $\log g$,
we used the MARCS model atmosphere (Castelli and Kurucz 2003), Synthe radiative
code (Kurucz 2005), solar abundances provided in Asplund \etal (2009), and VALD atomic
line list (Piskunov \etal 1995). Due to the low spectral resolution, our analysis was based
on the Balmer lines only (H$\alpha$, H$\beta$, and H$\gamma$). The parameters $T_{\rm eff}$
and $\log g$ were fitted with starting values of 7000 K and 4.0, respectively.
Microturbulence velocity was fixed at 2 km/s, while macroturbulence velocity at 4 km/s.
The rotational velocity $v \sin i$ was assumed to be of 5 km/s and the metallicity [M/H]
of 0.0 dex. Results of the analysis are collected in Table~2. In Fig.~3,
we present normalized spectra for three shortest-period $\delta$~Sct stars
in our sample with models obtained from the spectral synthesis.

\begin{table}[h]
\begin{center}
\caption{\small Atmospheric parameters of the observed stars}
\medskip
{\footnotesize
\begin{tabular}{ccc|ccc}
\hline
Name       & $T_{\rm eff}$ & $\log g$ & Name       & $T_{\rm eff}$ & $\log g$ \\
OGLE-      &       [K]     &          & OGLE-      &      [K]      &          \\
BLG-DSCT-  &               &          & BLG-DSCT-  &               &          \\
\hline
16832 & 7950 $\pm$ 260 & 3.69 $\pm$ 1.23 & 16844 & 7450 $\pm$ 220 & 4.31 $\pm$ 0.92 \\
16833 & 7610 $\pm$ 240 & 3.79 $\pm$ 1.03 & 16845 & 8170 $\pm$ 370 & 4.05 $\pm$ 1.05 \\
16834 & 7840 $\pm$ 190 & 4.68 $\pm$ 1.12 & 16846 & 8090 $\pm$ 310 & 4.73 $\pm$ 1.07 \\
16835 & 7570 $\pm$ 210 & 3.71 $\pm$ 0.64 & 16847 & 7770 $\pm$ 270 & 3.93 $\pm$ 0.83 \\
16836 & 8010 $\pm$ 220 & 4.35 $\pm$ 0.85 & 16848 & 8710 $\pm$ 360 & 4.08 $\pm$ 0.61 \\
16837 & 8230 $\pm$ 380 & 3.34 $\pm$ 0.58 & 16849 & 8130 $\pm$ 280 & 4.14 $\pm$ 0.91 \\
16838 & 7650 $\pm$ 300 & 4.90 $\pm$ 0.82 & 16850 & 8750 $\pm$ 500 & 4.47 $\pm$ 0.65 \\
16839 & 7730 $\pm$ 460 & 4.91 $\pm$ 0.80 & 16851 & 8930 $\pm$ 370 & 4.52 $\pm$ 0.39 \\
16840 & 8095 $\pm$ 183 & 4.97 $\pm$ 0.88 & 16852 & 7870 $\pm$ 300 & 4.08 $\pm$ 0.69 \\
16841 & 8410 $\pm$ 430 & 4.27 $\pm$ 1.06 & 16853 & 9480 $\pm$ 420 & 3.81 $\pm$ 0.62 \\
16842 & 8160 $\pm$ 290 & 4.59 $\pm$ 0.91 & 16854 & 7370 $\pm$ 230 & 4.34 $\pm$ 0.96 \\
16843 & 8390 $\pm$ 410 & 4.11 $\pm$ 0.83 & 16855 & 9180 $\pm$ 410 & 4.31 $\pm$ 0.68 \\
\hline
\end{tabular}}
\end{center}
\end{table}

\begin{figure}
\centerline{\includegraphics[angle=0,width=130mm]{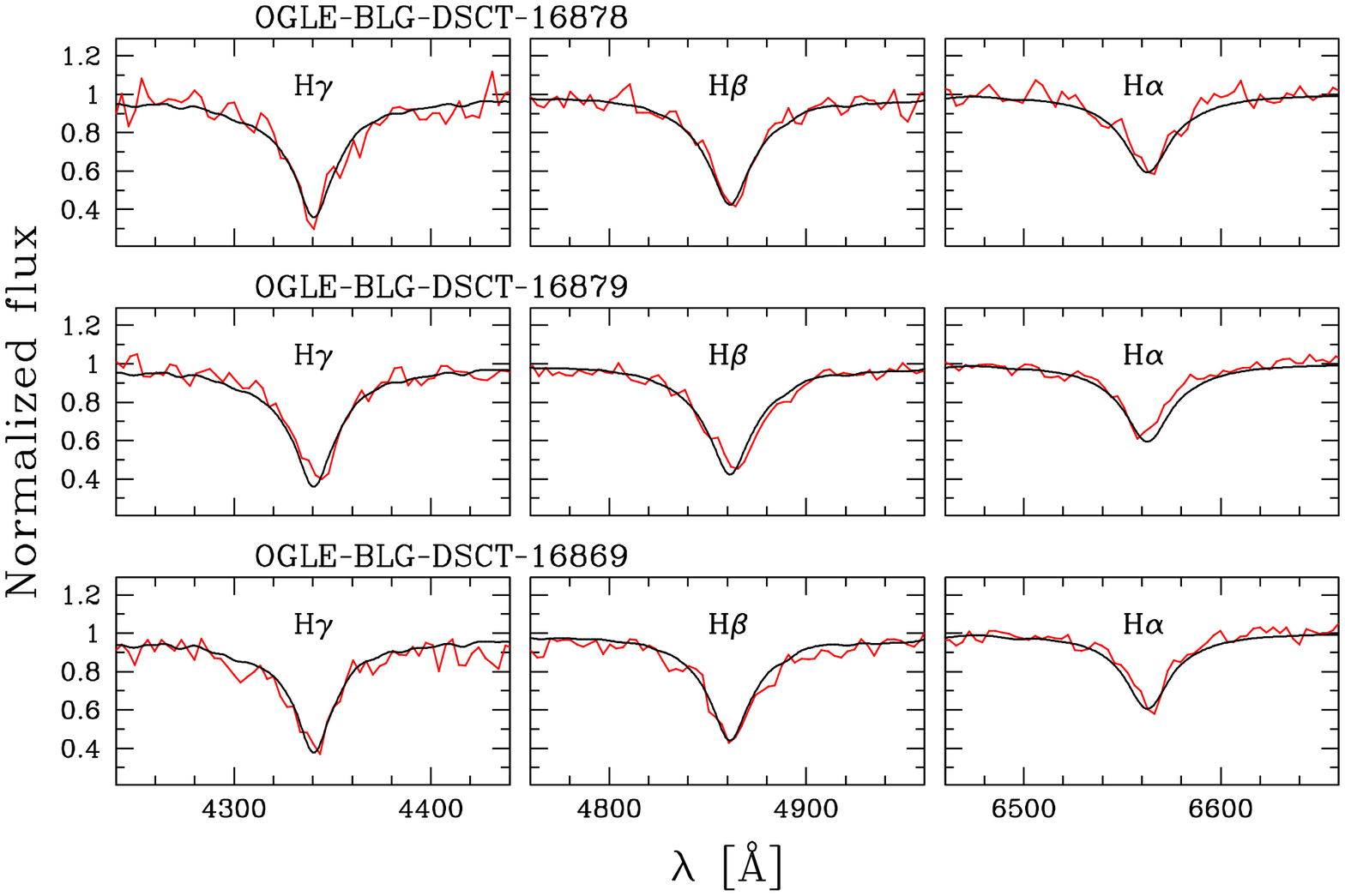}}
\FigCap{Fits to hydrogen lines in three $\delta$~Sct stars with periods below 30 min.}
\end{figure}

Fig.~4 shows the location of the 24 stars in the $T_{\rm eff}$ \vs $\log g$ plane.
All the objects have the atmospheric parameters typical for the main sequence A/F stars.
However, with the used spectral resolution, we cannot exclude that some objects
slightly evolved off the MS. Nevertheless, the spectra confirm that
all the stars are located in the area populated by $\delta$~Sct-type variables
in the Hertzsprung-Russell diagram.

We note that only one object, OGLE-BLG-DSCT-16883, was reported as
a variable source in Gaia DR3 (Gaia Collaboration 2021),
but it was incorrectly classified as a spotted star (of RS CVn type).
Measured parallaxes of 23 stars are between $0.20\pm0.06$ mas and $0.82\pm0.02$ mas,
which places them likely in the Galactic disk. The parallax of OGLE-BLG-DSCT-16864,
$0.08\pm0.06$ mas, indicates a more distant location, probably the Galactic bulge.

\begin{figure}[htb!]
\centerline{\includegraphics[angle=0,width=130mm]{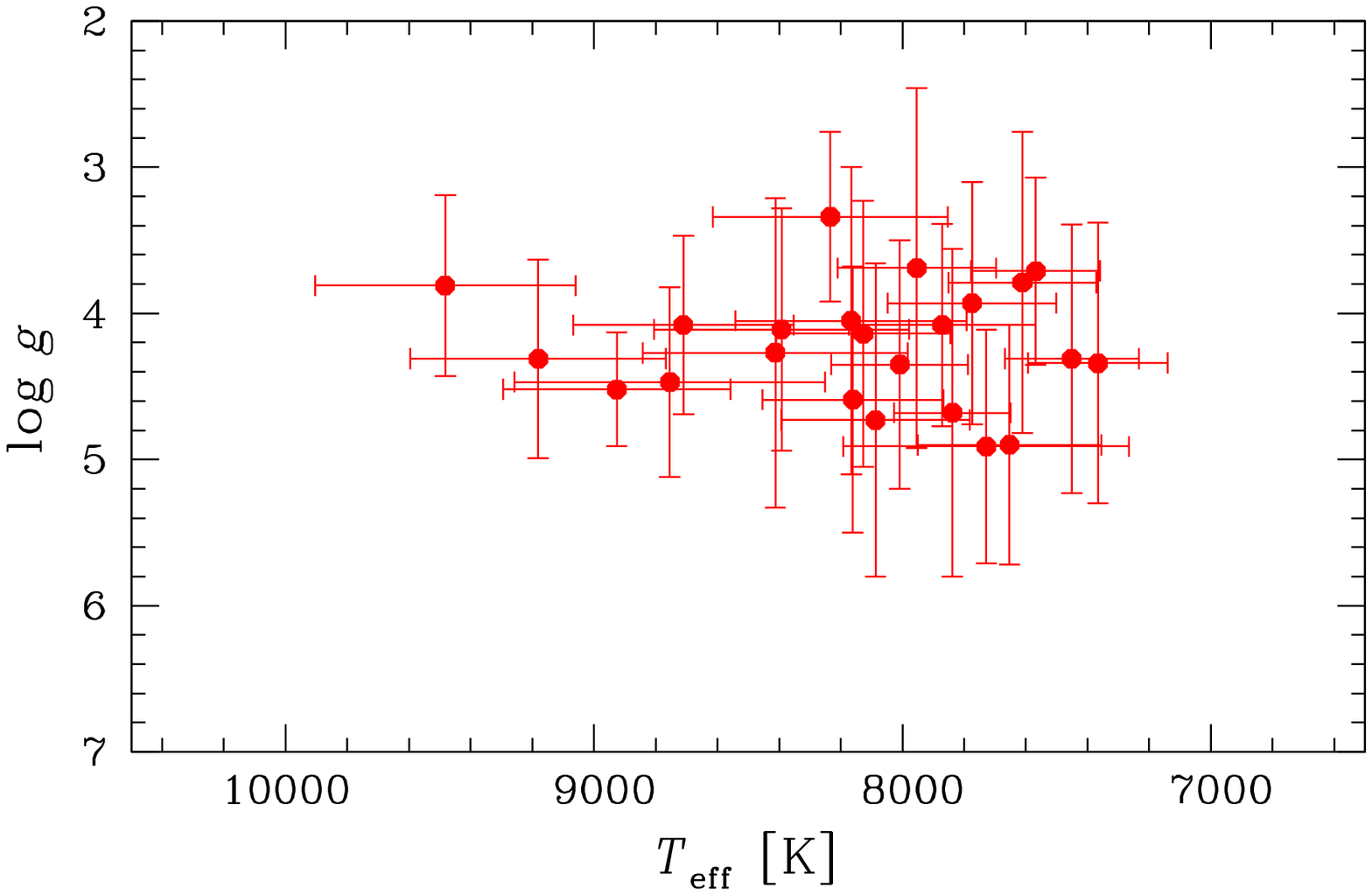}}
\FigCap{Positions of the 24 observed stars in the $T_{\rm eff}$ \vs $\log g$ plane. All
the objects are located in the area for MS stars of spectral type A and early F.}
\end{figure}


\Section{On-line Data}

The newly confirmed $\delta$~Sct stars supplement the OCVS.
The objects are arranged according to increasing right ascension and have
names from OGLE-BLG-DSCT-16860 to OGLE-BLG-DSCT-16883. The photometric
as well as spectroscopic data are available to the astronomical community
through the OGLE On-line Data Archive at {\it https://ogle.astrouw.edu.pl}.


\Acknow{
MG is supported by the EU Horizon 2020 research and innovation programme
under grant agreement No 101004719. This work is based on observations
collected at the European Southern Observatory under ESO programme 105.20EF.001.
We used data from the European Space Agency (ESA) mission Gaia,
processed by the Gaia Data Processing and Analysis Consortium (DPAC).
Funding for the DPAC has been provided by national institutions, in particular
the institutions participating in the Gaia Multilateral Agreement.
}


\end{document}